\documentclass[%
 aip,
rsi,%
 amsmath,amssymb,
 reprint,%
]{revtex4-1}

\usepackage{graphicx}
\usepackage{dcolumn}
\usepackage{bm}
\usepackage{mathtools}
\usepackage{mathpazo}
\usepackage{xcolor}
\usepackage{pgfplots}
\usepackage{caption,subcaption}

\begin{document}

\title[LN-CASS]{Simultaneous Parameter Estimation and Variable Selection via the LN-CASS Prior}

\author{W. Thomson}%
 \affiliation{School of Mathematics, University of Birmingham, UK}%
 \email{wmt095@bham.ac.uk}%

\author{S. Jabbari}%
\affiliation{School of Mathematics, University of Birmingham, UK}%
\affiliation{Institute of Microbiology \& Infection, University of Birmingham, UK}%

\author{A. E. Taylor}%
\affiliation{Institute of Microbiology \& Infection, University of Birmingham, UK}%
\affiliation{Centre for Endocrinology, Diabetes and Metabolism, Birmingham Health Partners, Birmingham, B15 2TT, United Kingdom
}%

\author{W. Arlt}%
\affiliation{Institute of Metabolism \& Systems Research, University of Birmingham, UK}%
\affiliation{Centre for Endocrinology, Diabetes and Metabolism, Birmingham Health Partners, Birmingham, B15 2TT, United Kingdom
}%

\author{D. J. Smith}%
\affiliation{School of Mathematics, University of Birmingham, UK}%
\affiliation{Institute of Metabolism \& Systems Research, University of Birmingham, UK}%
\affiliation{Centre for Endocrinology, Diabetes and Metabolism, Birmingham Health Partners, Birmingham, B15 2TT, United Kingdom
}%

\date{\today}
\maketitle

\begin{quotation}
We introduce a Bayesian prior distribution, the Logit-Normal continuous analogue of the spike-and-slab (LN-CASS), which enables flexible parameter estimation and variable/model selection in a variety of settings. We demonstrate its use and efficacy in three case studies -- a simulation study and two studies on real biological data from the fields of metabolomics and genomics. The prior allows the use of classical statistical models, which are easily interpretable and well-known to applied scientists, but performs comparably to common machine learning methods in terms of generalisability to previously unseen data.
\end{quotation}

\section{Introduction}

Often in real-world regression problems, we are faced with a situation in which we have a large number of potentially irrelevant predictors, possibly even greater than the number of observations. This so-called $p \gg n$ problem is especially prevalent in the biological and medical sciences with the advent of high-throughput experimental methods and an increasing focus on synthesising knowledge of molecular details into models predicting much lower-dimensional observable outcomes. Regularisation and shrinkage methods aim to reduce the influence of the inherent noise in such problems and provide sparse estimated parameter vectors, essentially performing simultaneous variable selection and parameter fitting. The motivation is two-fold. Firstly, regularisation aims to more robustly distinguish strong from weak effects, i.e.\ more reliably identify the genuine driving forces of the process of interest. Secondly, we wish to reduce overfitting to improve the generalisability of our models. The performance of our method in both of these respects is demonstrated below.

The most common means of dealing with the $p \gg n$ problem is the LASSO \cite{lasso,natmethlasso}, whose ability to induce genuine sparsity (i.e.\ estimates of \textit{exactly} zero) and whose computationally efficient implementation make it attractive for general-purpose regularised regression. A number of Bayesian analogues of the LASSO and other penalised likelihood methods have been proposed in order to more fully account for the uncertainty in parameter estimates, which we contend is particularly important in small $n$/large $p$ problems, and to tackle the tendency of the LASSO to underestimate large effects \cite{horseshoe,blasso,griffbrown}.

Some authors have focused on the subset of such problems in which the predictors have a known grouping structure, for example in problems from genetics in which the groups correspond to known regulatory networks \cite{gplassogene}. This has led to the development of both penalised likelihood \cite{sgl} and Bayesian \cite{bsgl,bgrplasso} modifications of common shrinkage methods.

In this paper, we present a new shrinkage prior -- the Logit-normal continuous analogue of the spike-and-slab (LN-CASS) -- based on a Logit-Normal relaxation of the Bernoulli distribution used in the spike-and-slab prior \cite{spikeslab}. The spike-and-slab is considered the gold-standard of Bayesian variable selection \cite{goldstandard}, but is computationally intractable in practice due to its combinatorial complexity.

The LN-CASS prior has the advantage that its intuitive formulation allows it to be simply extended to almost any hierarchical situation -- two of which are covered below -- allowing the modeller to tailor the specifications of common statistical models to favour `simpler' models in a variety of senses. Below we structure our models to favour first homogeneous groups of predictors before allowing within-group heterogeneity (simulation study) and to favour purely linear effects before nonlinear effects (metabolomics study), as well as applying the method in its simplest form to shrink logistic regression coefficients (microarray case study). The Bayesian formalism `allows the data to decide' the appropriate level of complexity through the likelihood function.

In the simulation study, the LN-CASS prior empirically appears robust to group misspecification, and outperforms the horseshoe prior \cite{horseshoe}, the LASSO \cite{lasso} and the sparse group lasso \cite{sgl}. Additionally, we apply the LN-CASS prior to a real-life classification task, in which we aim to distinguish benign from malignant adrenal tumours. Our method leads to an out-of-sample predictive performance comparable to state-of-the-art machine learning methods, but offers more interpretable results. We also use the method to build a predictive model of colon cancer malignancy using the well-known Colon dataset \cite{alon}. An implementation of the method for a variety of common statistical models is available in an \textsc{r} package (see Code Availability).

\begin{figure*}
	\begin{subfigure}{0.5\linewidth}
		\includegraphics[width = \linewidth]{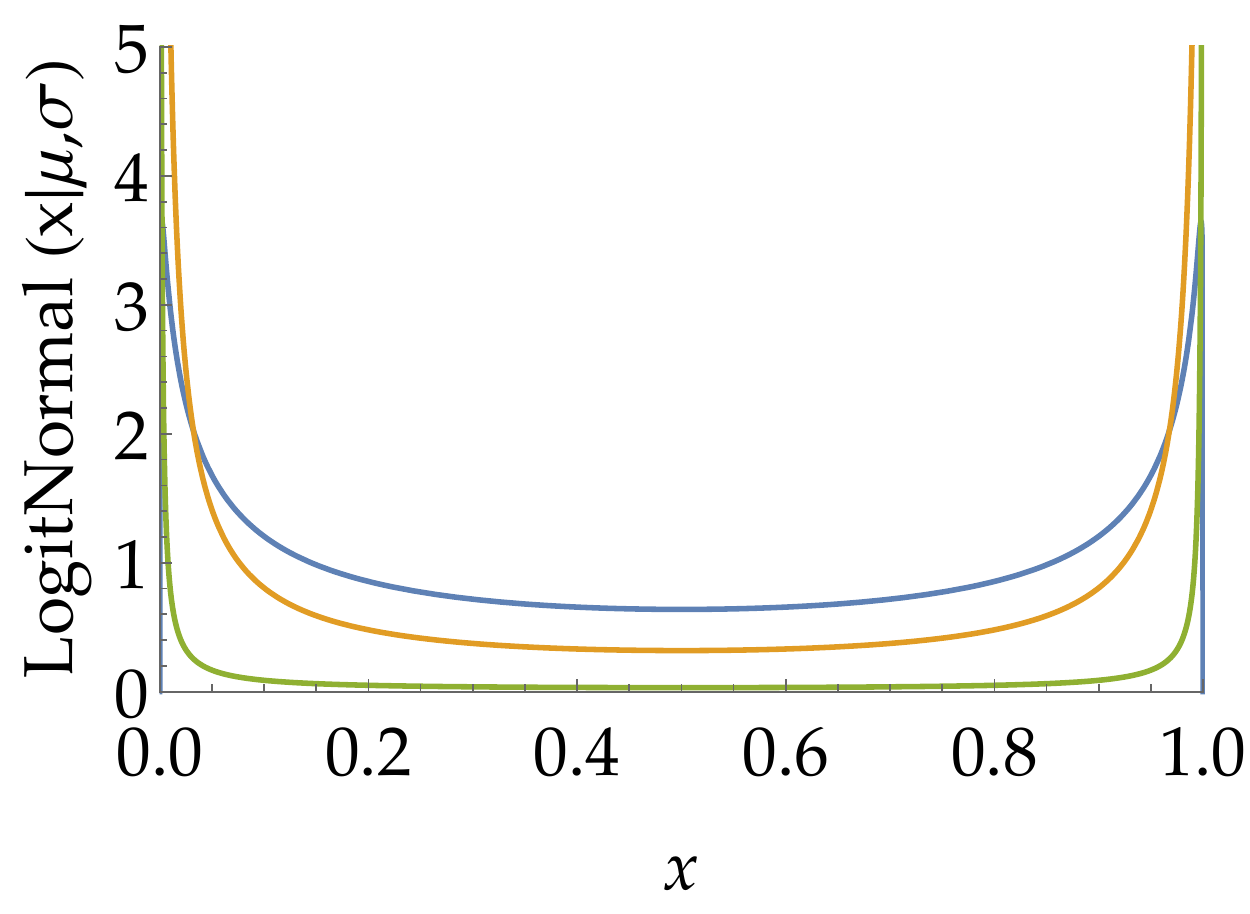}
		\caption{}
	\end{subfigure}%
	~
	\begin{subfigure}{0.5\linewidth}
		\includegraphics[width = \linewidth]{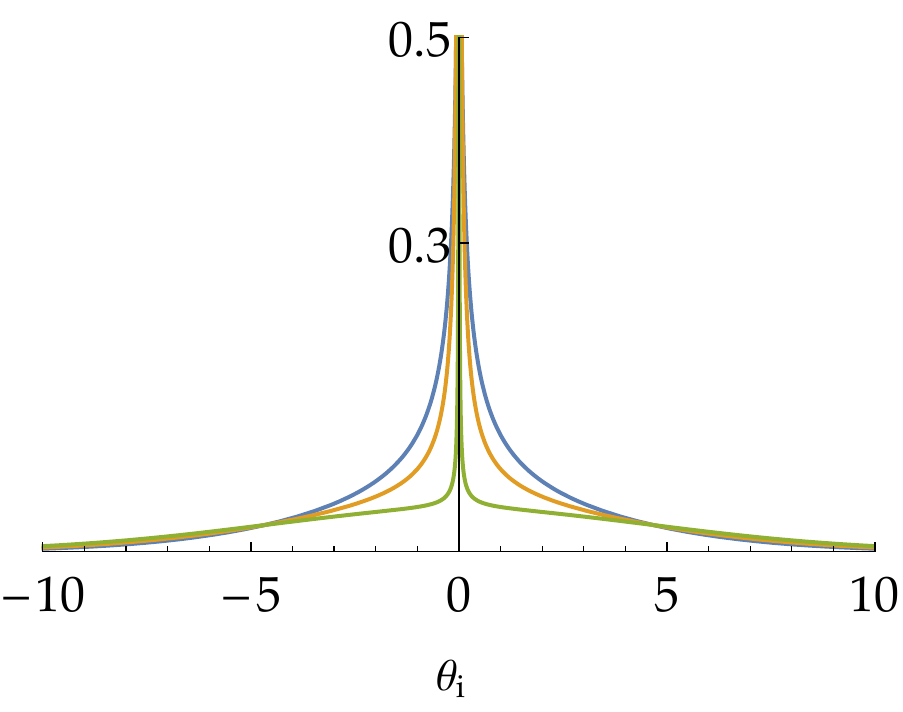}
		\caption{}
	\end{subfigure}
	\caption{(a) The Logit-Normal distribution with $\mu = 0$ and $\sigma$ given by (blue, orange, green) = $(2.5, 5, 50)$. (b) The LN-CASS priors induced by the Logit-Normal distributions of (a).}
	\label{prior}
\end{figure*}

\section{Results}

To illustrate the utility of the LN-CASS prior, we conducted three comparative studies with the intention of assessing its two primary functions: identification of genuinely non-zero effects and improving out-of-sample performance by reducing overfitting. Additionally, we chose two of the three settings to highlight the flexibility of the approach, in particular its capacity to include known group structure (simulation, section C.1) and to perform nonparametric regression (metabolomics, section C.2). These two extensions are by no means exhaustive, but are illustrative of the myriad possible areas of application (see Discussion).

\subsection{The LN-CASS Prior}

We now provide a brief outline of the LN-CASS prior. Mathematical details are available in the Online Methods section.

The fundamental motivation for developing the LN-CASS prior is to provide a computationally tractable alternative to the theoretical gold-standard of Bayesian variable selection, the spike-and-slab prior.

The spike-and-slab prior is based on the simple idea that, \textit{a priori}, we believe each parameter has some non-zero probability of being zero, and the rest of the probability mass is assigned to other plausible parameter values (often uniformly). This hard zero/non-zero distinction introduces a discrete component into our prior beliefs and renders the practical use of the prior combinatorially intractable -- we need to visit $2^p$ parameter combinations in order to adequately cover the parameter space where $p$ is the number of parameters in our model. For even moderately sized problems, this complexity renders the spike-and-slab impractical. Indeed, this combinatorial complexity is the same problem faced by the frequentist 'best subset selection', in which every possible subset of parameters is compared and the best performing subset is chosen.

By constructing a fully continuous approximation to the mixed spike-and-slab prior, we enable greatly improved sampling efficiency at the cost of relaxing the hard distinction between zero and non-zero parameters. The LN-CASS prior accomplishes this relaxation by replacing the discrete Bernoulli distribution in the mixture formulation of the spike-and-slab with a Logit-normal distribution (see Online Methods for details). The Logit-normal distribution, with suitable parameter choices, is a U-shaped distribution on $(0,1)$, assigning most of its mass to values close to the endpoints (figure \ref{prior}). The reason for choosing the Logit-Normal distribution for this purpose over the similar and more common Beta distribution is that it can be expressed as a transformation of standard normal random variables, which greatly aids the convergence properties of our sampler. Indeed, models can be specified purely in terms of parameters with (conditionally) standard normal prior distributions.

We interpret the values of the Logit-normal random variable as approximate variable inclusion probabilities, which allows simple propagation of these probabilities through a hierarchical prior structure. For example, in the simulation study below we impose a hierarchical prior structure in which we favour first exclusion of whole groups of variables, then allow inclusion of groups with a shared parameter, and finally allow groups with differing parameters. In the metabolomics case study below, we utilise this prior structure to favour linear effects first, before allowing non-linear effects if the data support such effects strongly enough. This corresponds to imposing a hierarchy on the complexity of the model and allows us to refine exactly how we control model complexity.

\subsection{Performance measures}

The main measure of performance we employ is the area under the Receiver Operating Characteristic (ROC) curve (AUC). The ROC curve is a plot of the false-positive rate (specificity) against the true-positive rate (sensitivity) as the probability threshold for classifying a prediction as positive or negative is varied. The AUC is interpretable as the probability of successfully distinguishing a positive result from a negative result, i.e.\ the probability of correctly assigning a larger predicted value to a positive case than a negative case. An AUC of 0.5 corresponds to a model that simply uses the class proportions as a prediction, while an AUC of 1 corresponds to a classifier which perfectly distinguishes positive and negative cases at some threshold. We use the AUC to quantify the trade-off between false- and true-positives in two settings. Firstly, in the simulation study the AUC is used to quantify the degree to which each method uncovers the correct ordering of ground-truth parameter values -- the degree to which genuinely small parameters are estimated to be small, and large parameters to be large. Secondly, in the metabolomics and microarray case studies, we use the AUC in the more conventional setting of quantifying the out-of-sample performance of a classifier. 

To quantify the agreement between the estimated and true parameters in the simulation study, we use the mean absolute error (MAE). The MAE is simply the average distance of the estimated from the true parameters.

\subsection{Applications}

\begin{figure*}[t]
	\begin{minipage}{0.5\linewidth}
		\centering
		\includegraphics[width = 0.85\linewidth]{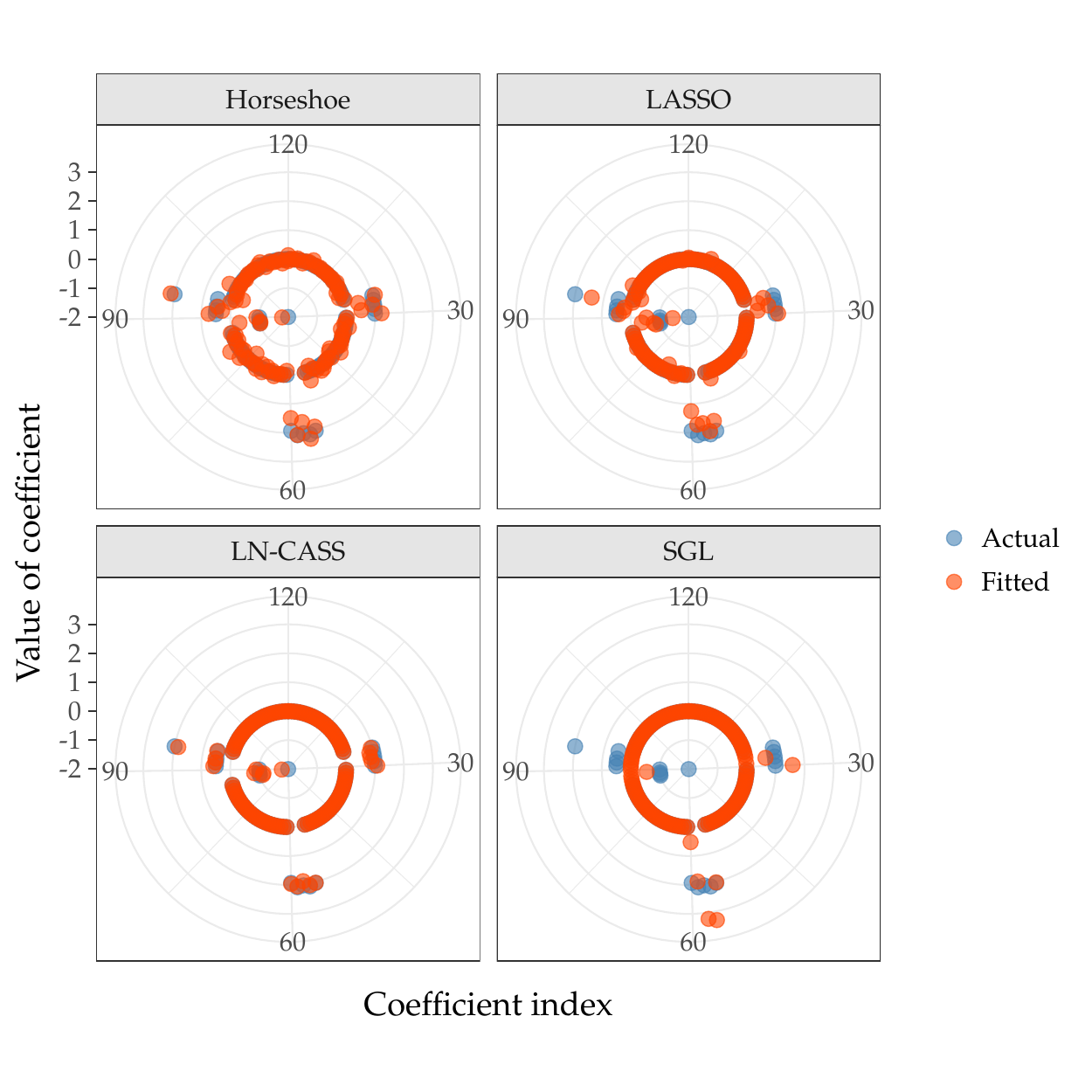}
		\vtop{(a)}
	\end{minipage}%
	\begin{minipage}{0.5\linewidth}
		\centering
		\hbox{\hspace{5ex} \vspace{5ex} \includegraphics[width = \linewidth, height = 0.5\linewidth]{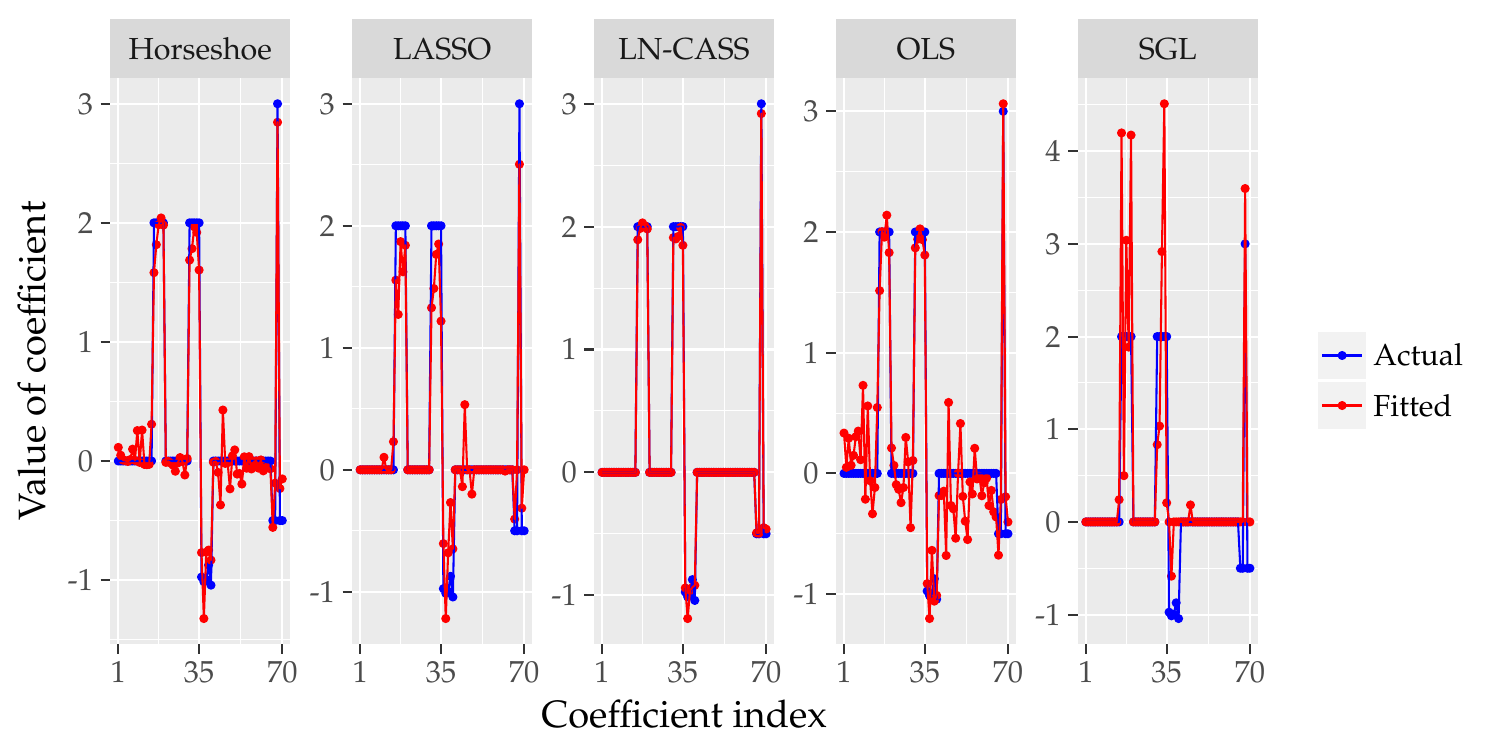}}
		\vspace{-5ex}
		\hbox{\hspace{0.5\linewidth} \vspace{2ex}(b)}
		\begin{tabular}{ccccccccccc}
			& \multicolumn{2}{c}{\textbf{HS}} & \multicolumn{2}{c}{\textbf{LASSO}} & \multicolumn{2}{c}{\textbf{LN-CASS}} & \multicolumn{2}{c}{\textbf{OLS}} & \multicolumn{2}{c}{\textbf{SGL}} \\ \hline
			& MAE            & AUC            & MAE             & AUC              & MAE                 & AUC            & MAE             & AUC            & MAE             & AUC            \\ \hline
			$p = 20$ & 0.196          & 1              & 0.223           & 0.99             & \textbf{0.043}      & 1              & 0.317           & 1              & 0.643           & 0.95           \\
			$p = 70$ & 0.111          & 0.98           & 0.13            & 0.941            & \textbf{0.017}      & \textbf{1}     & 0.22            & 0.94           & 0.31            & 0.791          \\
			$p =120$ & 0.096          & 0.9755         & 0.075           & 0.9702           & \textbf{0.015}      & \textbf{1}     & NA              & NA             & 0.145           & 0.7 \\
		\end{tabular}
		\vbox{\vspace{3ex} \hspace{3ex} (c)}
	\end{minipage}
	\caption{Agreement between ground truth and estimated parameters for the simulation study in the (a) $p = 120$ case (b) $p = 70$ case; (c) performance measures for each method.}
\end{figure*}

We now present the results of three case studies to evaluate the comparative ability of the LN-CASS prior to perform its two main duties -- sparse parameter estimation and improving out of sample performance. In the first case study, we attempt to recover ground-truth parameters in a simulation study in which we impose a known grouping structure in the predictors. In the second case study, we use real-world metabolomics data \cite{metabolomdata} to build a predictive model of adrenal tumour malignancy. In the third, we apply the LN-CASS prior in the context of Bayesian logistic regression to the well-known colon cancer dataset \cite{alon}.

\subsubsection{Simulation Study (grouped predictors)}

The motivation for this case study is to illustrate the ability of the LN-CASS prior to penalise not only model complexity in terms of the number of parameters, but also the granularity of the model. Such a formulation might be applied when there is some `subset' or `tree-like' structure in the predictors. For example, in immunological applications, cell subsets are often nested -- T-cells are subdivided into CD4$^+$ and CD8$^+$ T-cells, which in turn are subdivided into naive and memory subsets. The grouped LN-CASS prior favours within group homogeneity, essentially favouring less granular models, i.e.\ a model using total T-cell counts would be favoured over a model using CD4$^+$ and CD8$^+$ subsets as predictors.

We generated a simulated dataset of $n = 100$ observations from the linear regression model
\begin{equation}
	y_i = \beta_0 + \boldsymbol{X}_i \boldsymbol{\beta} + \varepsilon_i,
\end{equation}
for three different settings with grouped predictors, i.e.\ where pre-specified groups share mostly the same or similar parameter values (table 1, Online Methods). The matrix $\boldsymbol{X}$ was sampled from a Unit Latin Hypercube. The $\varepsilon_i$ were chosen to be i.i.d. zero-mean Gaussian. 

We then fit the model in \textsc{r} with the following methods for each of the three settings ($p = 20,70,120$): Group LN-CASS, LASSO, Horseshoe, Sparse Group LASSO and Ordinary Least Squares. Ordinary least squares was tested only for the $p<n$ cases because the problem is not well-defined when $p>n$. For code see Code Availability. Details of the models are available in Online Methods.

LN-CASS substantially outperforms all of the other methods in recovering the ground-truth parameters and correctly identifying zero parameters (figure 2).

\subsubsection{Steroid metabolomics and adrenal tumour malignancy (hierarchical GAM)}

We applied a hierarchical version of the LN-CASS prior to clinical data regarding the concentrations of metabolites in the urine of patients with two different adrenal tumours. The task was to predict the tumour type based on the metabolites, and to do this we used a generalised additive model (GAM) with logit link. The implementation of the prior in a hierarchical fashion here was strongly inspired by a recent paper by Griffin and Brown \cite{griffbrowngam}.

The GAM we implemented represented the effect of each covariate as the sum of linear basis functions. We imposed a hierarchy through the LN-CASS prior which favoured firstly the complete removal of a covariate from the model, then inclusion of a purely linear effect, and finally allowed each of the basis functions to be used to construct a non-linear effect (for details, see Online Methods).

The dataset consisted of 158 measurements of 32 covariates \cite{metabolomdata} collected as part of the EURINE-ACT study, with 45 positive cases (malignant adrenal tumours). All of the covariates are measurements of steroid concentrations in urine samples taken from each of the patients. There is a small proportion of missing data (up to 7\% of a covariate's measurements), which we imputed via the \texttt{mice()} function in \textsc{r} \cite{mice}. We then $\log(1+x)$ transformed all of the data because many of the predictors spanned several orders of magnitude. We subsequently scaled all covariates to lie in the interval $[0,1]$.

\begin{figure*}
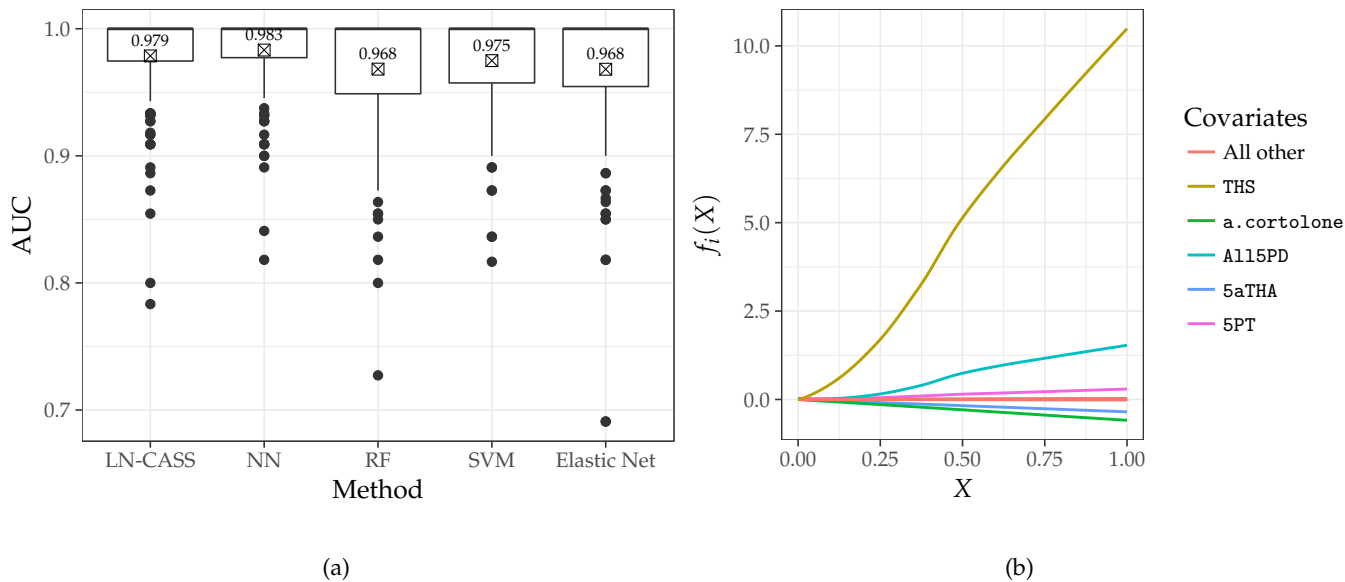

	\begin{subfigure}{0.5\linewidth}
	\input{metabolom_auc_cv.tikz}
	\caption{}
	\end{subfigure}%
~
	\begin{subfigure}{0.5\linewidth}
		\input{metabolom_splines.tikz}
		\caption{}
	\end{subfigure}
\caption{Metabolomics case study. (a) Boxplots of AUCs for each method computed via $16 \times 10$-fold cross-validation; (b) estimated mean functions $f_i$ from the LN-CASS hierarchical GAM. Functions have been smoothed for presentation purposes with a LOESS smoother using a small span.}
\label{metabolom}
\end{figure*}

We compared the classification performance of our hierarchical GAM with the performance of the following methods: Support Vector Machine (SVM), neural network (NN), random forest (RF) and elastic net (a modified version of the LASSO). Classification performance was measured using the mean AUC over $16 \times 10-$fold cross-validated runs. The results are presented in fig. \ref{metabolom} (a).

All of the methods perform comparably in terms of out-of-sample predictive performance, with the neural network performing the best and LN-CASS second in terms of both the mean and variability (inter-quartile range) of cross-validated AUCs. The authors are not aware of an appropriate and well-established statistical test to formalise the comparative performances of each method given the unequal variances, clear non-normality, and obvious dependency between samples for a given method. However, the Kruskal-Wallis test with a \textit{post hoc} Dunn test (and appropriate multiplicity correction) provides a non-parametric test for stochastic dominance (i.e.\ the tendency of values from one distribution to be larger than values from the other). We used two multiplicity corrections, both of which account for positive dependency (i.e.\ the tendency of large AUCs to be correlated within cross-validation folds). Using the Benjamini-Hochberg \cite{bhcorrection} correction, the only null hypotheses to be rejected at 95\% significance levels were that the distribution of AUCs for the neural network stochastically dominates those for the Elastic Net and the Random Forest ($p$-values 0.0344 and 0.0203, respectively). Using the Benjamini-Yekutieli \cite{bycorrection} correction, no null hypotheses were rejected; that is, no significant differences were found between the distributions in terms of stochastic dominance.

The results suggest that the out-of-sample performance of the hierarchical GAM with LN-CASS prior is comparable with that of state-of-the-art machine learning methods. We argue that this performance, in conjunction with the accuracy with which LN-CASS recovers `true' parameters and offers more classically interpretable results make it a valuable addition to the shrinkage and regularisation toolbox for applied scientists.

The recovered effects for each of the metabolites are presented in fig. \ref{metabolom} (b), as estimated from the full dataset. Clearly, the dominant predictor is \texttt{THS} which is in agreement with the original study, as are the influential roles of both \texttt{5PD} and \texttt{5PT}. We believe that the ability of the hierarchical GAM to produce plots such as these constitutes a considerable advantage over the machine learning methods tested and highlights the ability of LN-CASS to generate not only strong predictive models, but to be used as an exploratory tool for the generation of hypotheses for future study.

\subsection{Microarray data}

The final case study we conducted focused on the well-known Colon dataset of Alon et al. \cite{alon}. The dataset consists of measurements of the expression levels of 2000 genes in 62 subjects, with the response variable being an indicator of Colon cancer incidence, representing a typical $p \gg n$ problem in the biological/medical sciences. We compared the performance of logistic regression, with LN-CASS priors on the coefficients, to LASSO, Random Forest and Neural Network classifiers. We performed leave-one-out cross-validation (LOOCV) and computed the AUC across the left out samples in order to compare the estimated out-of-sample predictive accuracy of each method. In order to reduce the bias of the AUC estimates, we randomly removed an observation of the opposite class in each fold so that the class proportions were identical across folds.

We preprocessed the data by first log-transforming and subsequently standardising (i.e. subtracting the mean and dividing by the standard deviation) the expression level of each gene. We then screened the genes via a preliminary Wald test and selected the 500 genes with the largest $Z$-scores in absolute value, leaving us with a predictor matrix consisting of the expression levels of 500 genes in 62 tissues which acted as the input to all subsequent models.

The pooled LOOCV AUCs for each method were as follows: LN-CASS, 0.904; Neural Network, 0.8898; Random Forest, 0.8892; LASSO, 0.858. LN-CASS performs the best, but again all of the methods perform well and there is not a substantial difference between the estimated out-of-sample performance of each method.

\begin{figure}
	\input{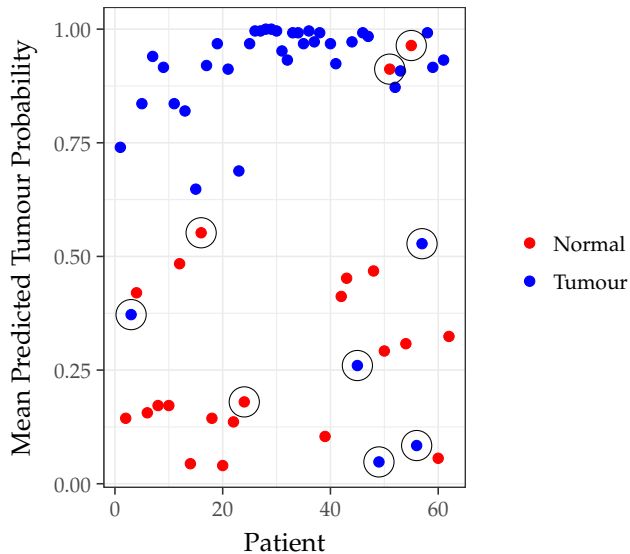}
	\caption{Mean predictions and observed outcomes from the LN-CASS model for the microarray data. Circled points have been identified as potentially mislabelled by \cite{alon,rslr}}
	\label{mislabelled}
\end{figure}

Interestingly, there is some biological evidence for class-mislabelling, i.e.\ samples being incorrectly marked as either tumour or healthy \cite{alon,rslr} in the colon dataset. According to Bootkrajang \& Kab{\'a}n \cite{rslr}, there are nine such samples. Figure \ref{mislabelled} shows the mean posterior prediction for each subject with these `suspicious' subjects circled. Clearly, there is reasonable agreement based on a visual inspection of the plot between the potentially mislabelled samples and those suggested by visual inspection of the LN-CASS model predictions. This suggests a possible secondary function of the LN-CASS prior in identifying mislabelled samples, the details however are left to future work.

\section{Discussion}

We have presented a new prior distribution for performing regularisation/shrinkage in a Bayesian framework. We have shown that its ability to produce generalisable predictive models is comparable to state-of-the art machine learning methods on two datasets of biological interest. Additionally, we have demonstrated with a simulation study the ability of our method to recover ground-truth parameters, even when the number of parameters is larger than the number of datapoints. In this regard, the performance of the LN-CASS prior is considerably better than other regularisation/shrinkage methods which aim to estimate the parameters of classical, generative probability models (linear regression, logistic regression etc.).

We believe that, combined, these two properties of the LN-CASS prior make it a worthwhile addition to the toolboxes of applied scientists working with typical biological datasets.

Our prior requires the choices of three hyperparameters, although we contend that they are much more interpretable than those required for other Bayesian shrinkage methods (see Online Methods). The three hyperparameters required correspond to, firstly, the standard deviation of the `slab' component, which we refer to as $\tau$; for standardised predictors, a default value of $\tau = 5$ has been sufficient for all of our applications because it essentially provides a vague Gaussian prior for non-zero coefficients. Secondly, the parameters of the Logit-Normal distribution (fig. \ref{prior} (a)) must be specified; we refer to these parameters as $\mu_\lambda$ and $\sigma_\lambda$. $\mu_\lambda$ can be chosen based on our prior beliefs about the probability of a zero coefficient, and in our experience does not require much tuning; the median of the Logit-Normal distribution is given by $\textrm{sigm}(\mu_\lambda)$, where $\textrm{sigm}(\cdot)$ is the logistic sigmoid function. Thus, if we believe \textit{a priori} that each coefficient has a probability $a$ of being non-zero, we simply set $\mu_\lambda = \textrm{logit}(a)$. $\sigma_\lambda$ simply controls the quality of the approximation to the spike-and-slab prior, with larger values corresponding to better approximations. We have used a default value of $\sigma_\lambda = 10$ throughout the paper; results are not sensitive to increases in this value.

The final key advantage of the LN-CASS prior is the ease with which it generalises to problems with a hierarchical complexity structure. This allows finer control of what exactly we mean by a `complex' model, and what we mean by a desirable model -- our example of using a generalised additive model for studying the metabolomics data above illustrates this point. In that case, we imposed a hierarchical complexity structure: no effect $\rightarrow$ linear effect $\rightarrow$ nonlinear effect. In the simulation study, we favoured a complexity structure: no effect $\rightarrow$ shared group effect $\rightarrow$ individual effect. These hierarchies are accomplished simply by propagating the value of the Logit-Normal random variable through each layer and taking its product with a new Logit-Normal random variable.

We note that the prior is particularly amenable to problems in which a hierarchical complexity structure is desired, by which we mean problems in which simpler models are nested within more complex models. The simplest case is the domain of the majority of the shrinkage/regularisation literature: models with fewer parameters are nested within models with more parameters. However, there are other problems with similar properties; linear models are nested within nonlinear models, models with some predictors sharing coefficients are nested within models in which each predictor has its own coefficient. One possible area of application is in multi-state survival modelling. Multi-state models describe transitions between disease states by distinct hazard functions, which may be difficult to fit with a small sample size. One might expect that the effects of many covariates remain fairly similar regardless of the state, for example age. Thus the LN-CASS prior could be used to introduce a `soft' constraint, encouraging but not enforcing covariates to share a parameter across hazard functions. This would essentially involve placing a grouped LN-CASS prior on the regression coefficients (as in the simulation study), with the groups corresponding to covariate effects.

As with most Bayesian methods, the main obstacle to the implementation of this methodology is the computational burden of MCMC sampling. Recent developments have made this procedure much more straightforward to implement and much faster \cite{rstan,nuts}. However, for large problems this computational burden is likely to be too large to compete with the much faster frequentist and machine learning methods available. Approximate Bayesian methods offer more computationally tractable alternatives to MCMC sampling, and would be an interesting avenue of future research for this problem and allow its scalability to very large problems. Any developments in this direction will be included in the available \textsc{r} package as and when they are made. In particular, nonparametric variational inference \cite{npvarinf} appears to be the most reasonable direction, since it is able to deal both with multimodal posterior distributions and non-conjugate prior distributions.

The LN-CASS method does not inherently provide `hard' variable selection, i.e.\ completely removing variables from the model, in the ilk of the LASSO. We advocate using the full model (i.e.\ including all predictors) for making predictions wherever possible, and using the absolute values of estimated parameters as variable importance measures for identifying the most important predictors for the purposes of hypothesis generation and/or obtaining biological insight. However, particularly in clinical/diagnostic circumstances, hard variable selection is useful to reduce the burden on clinicians/diagnosticians in collecting relevant data for utilising the model at the point of care.

A variety of applicable procedures for hard variable selection in Bayesian shrinkage models are available in an excellent review by Vehtari et al. \cite{vehtarireview}. One particularly simple method is to specify a threshold on the absolute values of the median parameters, i.e.\ discard all predictors whose absolute value is below some threshold. This threshold could be chosen based on the predictive performance of submodels containing only the predictors corresponding to the largest $k$ coefficients in absolute value -- one simply evaluates the predictive performance of each submodel and specifies a percentage of the maximum (i.e.\ the model including all variables) to retain.

To summarise, we have presented a flexible tool for performing regularised Bayesian regression in a variety of settings, which allows one to construct (with relative ease) problem-specific penalties on model complexity. The performance on out-of-sample data is typically at least as good as state-of-the-art methods, but the prior allows the use of classical statistical models which can be interpreted simply by applied biomedical scientists. \\

\textbf{Acknowledgements}

WT would like to thank the EPSRC for funding his doctoral research. SJ would like to acknowledge the BBSRC for funding (BB/M021386/1). SJ, DJS \& WT would like to thank the Wellcome Trust for funding the University of Birmingham's Model Parameterisation for Predictive Medicine workshop (1516ISSFFEL9).

\noindent \textbf{References}

\bibliography{natmethbib}

\begin{thebibliography}{22}%
\makeatletter
\providecommand \@ifxundefined [1]{%
 \@ifx{#1\undefined}
}%
\providecommand \@ifnum [1]{%
 \ifnum #1\expandafter \@firstoftwo
 \else \expandafter \@secondoftwo
 \fi
}%
\providecommand \@ifx [1]{%
 \ifx #1\expandafter \@firstoftwo
 \else \expandafter \@secondoftwo
 \fi
}%
\providecommand \natexlab [1]{#1}%
\providecommand \enquote  [1]{``#1''}%
\providecommand \bibnamefont  [1]{#1}%
\providecommand \bibfnamefont [1]{#1}%
\providecommand \citenamefont [1]{#1}%
\providecommand \href@noop [0]{\@secondoftwo}%
\providecommand \href [0]{\begingroup \@sanitize@url \@href}%
\providecommand \@href[1]{\@@startlink{#1}\@@href}%
\providecommand \@@href[1]{\endgroup#1\@@endlink}%
\providecommand \@sanitize@url [0]{\catcode `\\12\catcode `\$12\catcode
  `\&12\catcode `\#12\catcode `\^12\catcode `\_12\catcode `\%12\relax}%
\providecommand \@@startlink[1]{}%
\providecommand \@@endlink[0]{}%
\providecommand \url  [0]{\begingroup\@sanitize@url \@url }%
\providecommand \@url [1]{\endgroup\@href {#1}{\urlprefix }}%
\providecommand \urlprefix  [0]{URL }%
\providecommand \Eprint [0]{\href }%
\providecommand \doibase [0]{http://dx.doi.org/}%
\providecommand \selectlanguage [0]{\@gobble}%
\providecommand \bibinfo  [0]{\@secondoftwo}%
\providecommand \bibfield  [0]{\@secondoftwo}%
\providecommand \translation [1]{[#1]}%
\providecommand \BibitemOpen [0]{}%
\providecommand \bibitemStop [0]{}%
\providecommand \bibitemNoStop [0]{.\EOS\space}%
\providecommand \EOS [0]{\spacefactor3000\relax}%
\providecommand \BibitemShut  [1]{\csname bibitem#1\endcsname}%
\let\auto@bib@innerbib\@empty
\bibitem [{\citenamefont {Tibshirani}(1996)}]{lasso}%
  \BibitemOpen
  \bibfield  {author} {\bibinfo {author} {\bibfnamefont {R.}~\bibnamefont
  {Tibshirani}},\ }\href@noop {} {\bibfield  {journal} {\bibinfo  {journal}
  {Journal of the Royal Statistical Society. Series B (Methodological)}\
  }\textbf {\bibinfo {volume} {58}},\ \bibinfo {pages} {267} (\bibinfo {year}
  {1996})}\BibitemShut {NoStop}%
\bibitem [{\citenamefont {Lever}, \citenamefont {Krzywinski},\ and\
  \citenamefont {Altman}(2016)}]{natmethlasso}%
  \BibitemOpen
  \bibfield  {author} {\bibinfo {author} {\bibfnamefont {J.}~\bibnamefont
  {Lever}}, \bibinfo {author} {\bibfnamefont {M.}~\bibnamefont {Krzywinski}}, \
  and\ \bibinfo {author} {\bibfnamefont {N.}~\bibnamefont {Altman}},\
  }\href@noop {} {\bibfield  {journal} {\bibinfo  {journal} {Nature Methods}\
  }\textbf {\bibinfo {volume} {13}},\ \bibinfo {pages} {803} (\bibinfo {year}
  {2016})}\BibitemShut {NoStop}%
\bibitem [{\citenamefont {Carvalho}, \citenamefont {Polson},\ and\
  \citenamefont {Scott}(2010)}]{horseshoe}%
  \BibitemOpen
  \bibfield  {author} {\bibinfo {author} {\bibfnamefont {C.~M.}\ \bibnamefont
  {Carvalho}}, \bibinfo {author} {\bibfnamefont {N.~G.}\ \bibnamefont
  {Polson}}, \ and\ \bibinfo {author} {\bibfnamefont {J.~G.}\ \bibnamefont
  {Scott}},\ }\href@noop {} {\bibfield  {journal} {\bibinfo  {journal}
  {Biometrika}\ }\textbf {\bibinfo {volume} {97}},\ \bibinfo {pages} {465}
  (\bibinfo {year} {2010})}\BibitemShut {NoStop}%
\bibitem [{\citenamefont {Park}\ and\ \citenamefont {Casella}(2008)}]{blasso}%
  \BibitemOpen
  \bibfield  {author} {\bibinfo {author} {\bibfnamefont {T.}~\bibnamefont
  {Park}}\ and\ \bibinfo {author} {\bibfnamefont {G.}~\bibnamefont {Casella}},\
  }\href@noop {} {\bibfield  {journal} {\bibinfo  {journal} {Journal of the
  American Statistical Association}\ }\textbf {\bibinfo {volume} {103}},\
  \bibinfo {pages} {681} (\bibinfo {year} {2008})}\BibitemShut {NoStop}%
\bibitem [{\citenamefont {Griffin}\ and\ \citenamefont
  {Brown}(2010)}]{griffbrown}%
  \BibitemOpen
  \bibfield  {author} {\bibinfo {author} {\bibfnamefont {J.~E.}\ \bibnamefont
  {Griffin}}\ and\ \bibinfo {author} {\bibfnamefont {P.~J.}\ \bibnamefont
  {Brown}},\ }\href@noop {} {\bibfield  {journal} {\bibinfo  {journal}
  {Bayesian Analysis}\ }\textbf {\bibinfo {volume} {5}},\ \bibinfo {pages}
  {171} (\bibinfo {year} {2010})}\BibitemShut {NoStop}%
\bibitem [{\citenamefont {Jacob}, \citenamefont {Obozinski},\ and\
  \citenamefont {Vert}(2009)}]{gplassogene}%
  \BibitemOpen
  \bibfield  {author} {\bibinfo {author} {\bibfnamefont {L.}~\bibnamefont
  {Jacob}}, \bibinfo {author} {\bibfnamefont {G.}~\bibnamefont {Obozinski}}, \
  and\ \bibinfo {author} {\bibfnamefont {J.-P.}\ \bibnamefont {Vert}},\ }in\
  \href@noop {} {\emph {\bibinfo {booktitle} {{Proceedings of the 26th annual
  international conference on machine learning}}}}\ (\bibinfo {organization}
  {ACM},\ \bibinfo {year} {2009})\ pp.\ \bibinfo {pages} {433--440}\BibitemShut
  {NoStop}%
\bibitem [{\citenamefont {Simon}\ \emph {et~al.}(2013)\citenamefont {Simon},
  \citenamefont {Friedman}, \citenamefont {Hastie},\ and\ \citenamefont
  {Tibshirani}}]{sgl}%
  \BibitemOpen
  \bibfield  {author} {\bibinfo {author} {\bibfnamefont {N.}~\bibnamefont
  {Simon}}, \bibinfo {author} {\bibfnamefont {J.}~\bibnamefont {Friedman}},
  \bibinfo {author} {\bibfnamefont {T.}~\bibnamefont {Hastie}}, \ and\ \bibinfo
  {author} {\bibfnamefont {R.}~\bibnamefont {Tibshirani}},\ }\href@noop {}
  {\bibfield  {journal} {\bibinfo  {journal} {Journal of Computational and
  Graphical Statistics}\ }\textbf {\bibinfo {volume} {22}},\ \bibinfo {pages}
  {231} (\bibinfo {year} {2013})}\BibitemShut {NoStop}%
\bibitem [{\citenamefont {Chen}\ \emph {et~al.}(2016)\citenamefont {Chen},
  \citenamefont {Chu}, \citenamefont {Yuan},\ and\ \citenamefont {Wu}}]{bsgl}%
  \BibitemOpen
  \bibfield  {author} {\bibinfo {author} {\bibfnamefont {R.-B.}\ \bibnamefont
  {Chen}}, \bibinfo {author} {\bibfnamefont {C.-H.}\ \bibnamefont {Chu}},
  \bibinfo {author} {\bibfnamefont {S.}~\bibnamefont {Yuan}}, \ and\ \bibinfo
  {author} {\bibfnamefont {Y.~N.}\ \bibnamefont {Wu}},\ }\href@noop {}
  {\bibfield  {journal} {\bibinfo  {journal} {Journal of Computational and
  Graphical Statistics}\ }\textbf {\bibinfo {volume} {25}},\ \bibinfo {pages}
  {665} (\bibinfo {year} {2016})}\BibitemShut {NoStop}%
\bibitem [{\citenamefont {Xu}, \citenamefont {Ghosh}\ \emph
  {et~al.}(2015)\citenamefont {Xu}, \citenamefont {Ghosh} \emph
  {et~al.}}]{bgrplasso}%
  \BibitemOpen
  \bibfield  {author} {\bibinfo {author} {\bibfnamefont {X.}~\bibnamefont
  {Xu}}, \bibinfo {author} {\bibfnamefont {M.}~\bibnamefont {Ghosh}},  \emph
  {et~al.},\ }\href@noop {} {\bibfield  {journal} {\bibinfo  {journal}
  {Bayesian Analysis}\ }\textbf {\bibinfo {volume} {10}},\ \bibinfo {pages}
  {909} (\bibinfo {year} {2015})}\BibitemShut {NoStop}%
\bibitem [{\citenamefont {George}\ and\ \citenamefont
  {McCulloch}(1993)}]{spikeslab}%
  \BibitemOpen
  \bibfield  {author} {\bibinfo {author} {\bibfnamefont {E.~I.}\ \bibnamefont
  {George}}\ and\ \bibinfo {author} {\bibfnamefont {R.~E.}\ \bibnamefont
  {McCulloch}},\ }\href@noop {} {\bibfield  {journal} {\bibinfo  {journal}
  {Journal of the American Statistical Association}\ }\textbf {\bibinfo
  {volume} {88}},\ \bibinfo {pages} {881} (\bibinfo {year} {1993})}\BibitemShut
  {NoStop}%
\bibitem [{\citenamefont {Polson}\ and\ \citenamefont
  {Scott}(2010)}]{goldstandard}%
  \BibitemOpen
  \bibfield  {author} {\bibinfo {author} {\bibfnamefont {N.~G.}\ \bibnamefont
  {Polson}}\ and\ \bibinfo {author} {\bibfnamefont {J.~G.}\ \bibnamefont
  {Scott}},\ }\href@noop {} {\bibfield  {journal} {\bibinfo  {journal}
  {Bayesian Statistics}\ }\textbf {\bibinfo {volume} {9}},\ \bibinfo {pages}
  {501} (\bibinfo {year} {2010})}\BibitemShut {NoStop}%
\bibitem [{\citenamefont {Alon}\ \emph {et~al.}(1999)\citenamefont {Alon},
  \citenamefont {Barkai}, \citenamefont {Notterman}, \citenamefont {Gish},
  \citenamefont {Ybarra}, \citenamefont {Mack},\ and\ \citenamefont
  {Levine}}]{alon}%
  \BibitemOpen
  \bibfield  {author} {\bibinfo {author} {\bibfnamefont {U.}~\bibnamefont
  {Alon}}, \bibinfo {author} {\bibfnamefont {N.}~\bibnamefont {Barkai}},
  \bibinfo {author} {\bibfnamefont {D.~A.}\ \bibnamefont {Notterman}}, \bibinfo
  {author} {\bibfnamefont {K.}~\bibnamefont {Gish}}, \bibinfo {author}
  {\bibfnamefont {S.}~\bibnamefont {Ybarra}}, \bibinfo {author} {\bibfnamefont
  {D.}~\bibnamefont {Mack}}, \ and\ \bibinfo {author} {\bibfnamefont {A.~J.}\
  \bibnamefont {Levine}},\ }\href@noop {} {\bibfield  {journal} {\bibinfo
  {journal} {Proceedings of the National Academy of Sciences}\ }\textbf
  {\bibinfo {volume} {96}},\ \bibinfo {pages} {6745} (\bibinfo {year}
  {1999})}\BibitemShut {NoStop}%
\bibitem [{\citenamefont {Arlt}\ \emph {et~al.}(2011)\citenamefont {Arlt},
  \citenamefont {Biehl}, \citenamefont {Taylor}, \citenamefont {Hahner},
  \citenamefont {Libe}, \citenamefont {Hughes}, \citenamefont {Schneider},
  \citenamefont {Smith}, \citenamefont {Stiekema}, \citenamefont {Krone} \emph
  {et~al.}}]{metabolomdata}%
  \BibitemOpen
  \bibfield  {author} {\bibinfo {author} {\bibfnamefont {W.}~\bibnamefont
  {Arlt}}, \bibinfo {author} {\bibfnamefont {M.}~\bibnamefont {Biehl}},
  \bibinfo {author} {\bibfnamefont {A.~E.}\ \bibnamefont {Taylor}}, \bibinfo
  {author} {\bibfnamefont {S.}~\bibnamefont {Hahner}}, \bibinfo {author}
  {\bibfnamefont {R.}~\bibnamefont {Libe}}, \bibinfo {author} {\bibfnamefont
  {B.~A.}\ \bibnamefont {Hughes}}, \bibinfo {author} {\bibfnamefont
  {P.}~\bibnamefont {Schneider}}, \bibinfo {author} {\bibfnamefont {D.~J.}\
  \bibnamefont {Smith}}, \bibinfo {author} {\bibfnamefont {H.}~\bibnamefont
  {Stiekema}}, \bibinfo {author} {\bibfnamefont {N.}~\bibnamefont {Krone}},
  \emph {et~al.},\ }\href@noop {} {\bibfield  {journal} {\bibinfo  {journal}
  {The Journal of Clinical Endocrinology \& Metabolism}\ }\textbf {\bibinfo
  {volume} {96}},\ \bibinfo {pages} {3775} (\bibinfo {year}
  {2011})}\BibitemShut {NoStop}%
\bibitem [{\citenamefont {Griffin}, \citenamefont {Brown}\ \emph
  {et~al.}(2017)\citenamefont {Griffin}, \citenamefont {Brown} \emph
  {et~al.}}]{griffbrowngam}%
  \BibitemOpen
  \bibfield  {author} {\bibinfo {author} {\bibfnamefont {J.}~\bibnamefont
  {Griffin}}, \bibinfo {author} {\bibfnamefont {P.}~\bibnamefont {Brown}},
  \emph {et~al.},\ }\href@noop {} {\bibfield  {journal} {\bibinfo  {journal}
  {Bayesian Analysis}\ }\textbf {\bibinfo {volume} {12}},\ \bibinfo {pages}
  {135} (\bibinfo {year} {2017})}\BibitemShut {NoStop}%
\bibitem [{\citenamefont {Buuren}\ and\ \citenamefont
  {Groothuis-Oudshoorn}(2011)}]{mice}%
  \BibitemOpen
  \bibfield  {author} {\bibinfo {author} {\bibfnamefont {S.}~\bibnamefont
  {Buuren}}\ and\ \bibinfo {author} {\bibfnamefont {K.}~\bibnamefont
  {Groothuis-Oudshoorn}},\ }\href@noop {} {\bibfield  {journal} {\bibinfo
  {journal} {Journal of statistical software}\ }\textbf {\bibinfo {volume}
  {45}} (\bibinfo {year} {2011})}\BibitemShut {NoStop}%
\bibitem [{\citenamefont {Benjamini}\ and\ \citenamefont
  {Hochberg}(1995)}]{bhcorrection}%
  \BibitemOpen
  \bibfield  {author} {\bibinfo {author} {\bibfnamefont {Y.}~\bibnamefont
  {Benjamini}}\ and\ \bibinfo {author} {\bibfnamefont {Y.}~\bibnamefont
  {Hochberg}},\ }\href@noop {} {\bibfield  {journal} {\bibinfo  {journal}
  {Journal of the royal statistical society. Series B (Methodological)}\ ,\
  \bibinfo {pages} {289}} (\bibinfo {year} {1995})}\BibitemShut {NoStop}%
\bibitem [{\citenamefont {Benjamini}\ and\ \citenamefont
  {Yekutieli}(2001)}]{bycorrection}%
  \BibitemOpen
  \bibfield  {author} {\bibinfo {author} {\bibfnamefont {Y.}~\bibnamefont
  {Benjamini}}\ and\ \bibinfo {author} {\bibfnamefont {D.}~\bibnamefont
  {Yekutieli}},\ }\href@noop {} {\bibfield  {journal} {\bibinfo  {journal}
  {Annals of statistics}\ ,\ \bibinfo {pages} {1165}} (\bibinfo {year}
  {2001})}\BibitemShut {NoStop}%
\bibitem [{\citenamefont {Bootkrajang}\ and\ \citenamefont
  {Kab{\'a}n}(2013)}]{rslr}%
  \BibitemOpen
  \bibfield  {author} {\bibinfo {author} {\bibfnamefont {J.}~\bibnamefont
  {Bootkrajang}}\ and\ \bibinfo {author} {\bibfnamefont {A.}~\bibnamefont
  {Kab{\'a}n}},\ }\href@noop {} {\bibfield  {journal} {\bibinfo  {journal}
  {Bioinformatics}\ }\textbf {\bibinfo {volume} {29}},\ \bibinfo {pages} {870}
  (\bibinfo {year} {2013})}\BibitemShut {NoStop}%
\bibitem [{\citenamefont {{Stan Development Team}}(2016)}]{rstan}%
  \BibitemOpen
  \bibfield  {author} {\bibinfo {author} {\bibnamefont {{Stan Development
  Team}}},\ }\href {http://mc-stan.org/} {\enquote {\bibinfo {title} {{RStan}:
  the {R} interface to {Stan}},}\ } (\bibinfo {year} {2016}),\ \bibinfo {note}
  {r package version 2.14.1}\BibitemShut {NoStop}%
\bibitem [{\citenamefont {Hoffman}\ and\ \citenamefont {Gelman}(2014)}]{nuts}%
  \BibitemOpen
  \bibfield  {author} {\bibinfo {author} {\bibfnamefont {M.~D.}\ \bibnamefont
  {Hoffman}}\ and\ \bibinfo {author} {\bibfnamefont {A.}~\bibnamefont
  {Gelman}},\ }\href@noop {} {\bibfield  {journal} {\bibinfo  {journal}
  {Journal of Machine Learning Research}\ }\textbf {\bibinfo {volume} {15}},\
  \bibinfo {pages} {1593} (\bibinfo {year} {2014})}\BibitemShut {NoStop}%
\bibitem [{\citenamefont {Gershman}, \citenamefont {Hoffman},\ and\
  \citenamefont {Blei}(2012)}]{npvarinf}%
  \BibitemOpen
  \bibfield  {author} {\bibinfo {author} {\bibfnamefont {S.~J.}\ \bibnamefont
  {Gershman}}, \bibinfo {author} {\bibfnamefont {M.~D.}\ \bibnamefont
  {Hoffman}}, \ and\ \bibinfo {author} {\bibfnamefont {D.~M.}\ \bibnamefont
  {Blei}},\ }in\ \href@noop {} {\emph {\bibinfo {booktitle} {Proceedings of the
  29th International Coference on International Conference on Machine
  Learning}}},\ \bibinfo {series and number} {ICML'12}\ (\bibinfo {year}
  {2012})\BibitemShut {NoStop}%
\bibitem [{\citenamefont {Vehtari}, \citenamefont {Ojanen}\ \emph
  {et~al.}(2012)\citenamefont {Vehtari}, \citenamefont {Ojanen} \emph
  {et~al.}}]{vehtarireview}%
  \BibitemOpen
  \bibfield  {author} {\bibinfo {author} {\bibfnamefont {A.}~\bibnamefont
  {Vehtari}}, \bibinfo {author} {\bibfnamefont {J.}~\bibnamefont {Ojanen}},
  \emph {et~al.},\ }\href@noop {} {\bibfield  {journal} {\bibinfo  {journal}
  {Statistics Surveys}\ }\textbf {\bibinfo {volume} {6}},\ \bibinfo {pages}
  {142} (\bibinfo {year} {2012})}\BibitemShut {NoStop}%
\end{thebibliography}%

\section{Online Methods}

\subsection{The Basic LN-CASS prior (Section II.C.3, Microarray Data)}

The basic LN-CASS prior, as placed on the regression coefficients in the microarray case study in the main text, is used to shrink a single parameter in a non-hierarchical way, i.e.\ independently of the sizes of other parameters. It is the building block for the hierarchical complexity models used in the main text in the simulation and metabolomics case studies.

The marginal prior for a single regression coefficient can be specified as a scale mixture of normal densities (ref andrews, mallows) as follows,
\begin{align}
\theta_i|\tau,\lambda_i &\sim \mathcal{N}(0,(\lambda_i\tau)^2), \label{basic1}\\
\lambda_i &\sim \textrm{LogitNormal}(\mu_\lambda,\sigma_\lambda).\label{basic2}
\end{align}
All experiments in the main text used fixed hyperparameters $\tau,\mu_\lambda,\sigma_\lambda$ (see there for a discussion), but of course they could be given their own hyper-prior distributions. Additionally, in the main text, the hyperparameters $\mu_\lambda, \sigma_\lambda$ are equal for each $i$. It is straightforward to allow different hyperparameters for each covariate to encode prior beliefs about the inclusion probabilities of the individual covariates (see main text).

The mixture parameter $\lambda_i$ is analogous to an inclusion indicator for the variable of interest. Indeed, replacing the Logit-Normal prior with a Bernoulli prior yields a spike-and-slab model with a spike at $0$ and a normal slab with variance $\tau^2$, corresponding to a situation in which the $i^{th}$ variable is either completely excluded, or included with a $\mathcal{N}(0,\tau^2)$ prior.

Equations \eqref{basic1} \& \eqref{basic2} can be rewritten in the following way, yielding a model in which inference is performed on a new parameter vector $(\boldsymbol{\theta},\boldsymbol{\tilde{\lambda}})$ specified solely in terms of a multivariate normal prior with diagonal covariance matrix,
\begin{align*}
\theta_i|\tau,\lambda_i &\sim \mathcal{N}(0,(\lambda_i\tau)^2), \\
\tilde{\lambda}_i &\sim \mathcal{N}(\mu_\lambda,\sigma_\lambda^2), \\
\lambda_i &= \textrm{logit}^{-1}(\tilde{\lambda}_i).
\end{align*}

This formulation, empirically, greatly enhances the performance of the No-U-Turn Markov Chain Monte Carlo sampler (NUTS) used for making posterior inference and results in much improved convergence properties over similar priors (e.g.\ the horseshoe).

\subsection{Introducing Hierarchical Complexity Structure}

The interpretation of the parameters $\lambda_i$ as relaxed variable inclusion probabilities provides a simple way to impose soft hierarchical complexity constraints on models, by propagating these probabilities through the prior hierarchy.

The priors used for the simulation and metabolomics case studies in the main text use this structure to essentially define priors over models, with simplicity of models (in a context appropriate sense) favoured by the prior.

\subsubsection{The grouped LN-CASS prior (Section II.C.1), Simulation Study}

To wit, the grouped LN-CASS prior used in the simulation study is formulated as follows.

We assume that the coefficient for each covariate in a linear regression is composed of the sum of a group-level and a covariate-level coefficient, so that each covariate $X_i$ is associated with a parameter
\begin{equation*}
\beta_i = \theta_{G_i} + \theta_i
\end{equation*}
where $G_i$ is the (pre-specified, perhaps by clustering) group to which covariate $i$ belongs. Thus the effect of each covariate is due to a common effect, $\theta_{G_i}$, among all members of its group and a deviation, $\theta_i$, from the shared effect which is particular to that covariate.

The prior is constructed to favour exclusion of the whole group from the model, followed by inclusion of the group with a shared effect, followed by possibly distinct effects for each group element. This structure is accomplished by propagating the inclusion probabilities through the following prior structure,
\begin{align}
\theta_{G_i}|\lambda_{G_i} & \sim \mathcal{N}(0,(\lambda_{G_i}\tau)^2), \label{gptheta}\\
\lambda_{G_i} & \sim \textrm{LogitNormal}(\mu_\lambda,\sigma_\lambda), \label{gplambda} \\
\theta_i|\lambda_{G_i},\lambda_{i} & \sim \mathcal{N}(0,(\lambda_{G_i}\lambda_i\tau)^2), \label{indtheta}\\
\lambda_{i} & \sim \textrm{LogitNormal}(\mu_\lambda,\sigma_\lambda), \label{indlambda}.
\end{align}

Equations \eqref{gptheta}, \eqref{gplambda} specify an LN-CASS prior of the form \eqref{basic1}--\eqref{basic2} on the group level components of the parameters $\beta_i$, while equations \eqref{indtheta}, \eqref{indlambda} specify conditional LN-CASS priors on the individual covariate level components, with the conditioning being on the inclusion probabilities for the groups, i.e.\ given a small inclusion probability for the group, the individual level components are instantly assigned a small inclusion probability. However, given a group inclusion probability close to $1$, the individual components are essentially assigned their own LN-CASS priors of the form \eqref{basic1}, \eqref{basic2}, favouring a situation in which the group share the group-level component only.

\subsubsection{The hierarchical GAM LN-CASS prior (Section II.C.2, Metabolomics Data)}

For the metabolomics case study in the main text, the LN-CASS prior was used as part of a hierarchical generalised additive model (GAM). The set up is a logistic regression problem in which we suspect that some covariates may have nonlinear effects, but we wish to let the data decide whether including such effects is worthwhile for the purposes of prediction.

\begin{table*}[t]
	\begin{tabular}{lllll}
		& \textbf{Zero groups} & \textbf{Constant groups} & \textbf{Noisy groups} & \textbf{Disparate groups}\\
		\hline
		
		$p = 20$
		
		& \multicolumn{1}{l}{$\boldsymbol{\beta}_{G_1}, \boldsymbol{\beta}_{G_3} = \boldsymbol{0}$}
		
		& \multicolumn{1}{l}{$\boldsymbol{\beta}_{G_2} = \boldsymbol{2}$}	& \multicolumn{1}{l}{$\boldsymbol{\beta}_{G_4} = \boldsymbol{-1} + \mathcal{N}(0,0.1^2)$}	& \\

		$p = 70$                                                         
		
		& \begin{tabular}[c]{@{}l@{}} 10 zero groups total.
		\end{tabular}
		
		&  \multicolumn{1}{l}{$\boldsymbol{\beta}_{G_4},\boldsymbol{\beta}_{G_7} = \boldsymbol{2}$}		& \multicolumn{1}{l}{$\boldsymbol{\beta}_{G_8} = \boldsymbol{-1} + \mathcal{N}(0,0.1^2)$} & \multicolumn{1}{l}{$\boldsymbol{\beta}_{G_{14}} = (-0.5,-0.5,3,-0.5,-0.5)$}\\

		$p =120$       
		
		& \begin{tabular}[c]{@{}l@{}} 20 zero groups total.
		\end{tabular}

		& \multicolumn{1}{l}{$\boldsymbol{\beta}_{G_6} = \boldsymbol{1}$}  		& \multicolumn{1}{l}{$\boldsymbol{\beta}_{G_{12}} = \boldsymbol{2} + \mathcal{N}(0,0.1^2)$}	& \begin{tabular}[c]{@{}l@{}} $\boldsymbol{\beta}_{G_{18}} = (-1, -1, -1, -2, -1)$\\ $\boldsymbol{\beta}_{G_{19}} = (0.5, 0.5, 0.5, 2, 0.5)$
		\end{tabular}
	\end{tabular}
	\caption{\label{simstudytable} True coefficients for each of the simulation experiments. Bold numbers represent constant vectors of length 5 and we abuse the $\mathcal{N}(0,0.1^2)$ notation to mean a vector of 5 samples from a normal random number generator with mean 0 and standard deviation 0.1.}
\end{table*}

Generalised additive models are extensions of generalised linear models in which the linear predictor is replaced with a sum of nonlinear covariate effects, i.e.\
\begin{align}
g^{-1}(y_i) = f_0 + \sum_{i = 1}^{p} f_i(x_i),
\end{align}
with $g$ a link function, $f_0$ a constant, and the $f_i$ functions to be learnt.

We model the $f_i$ as piecewise linear functions with some pre-specified number of knots, $M$. Consider, without loss of generality, the covariate space $\chi = [0,1]^p$. The functions
\begin{align}
\varphi_k(x) = \begin{cases}
0, & x \leq x_k \\
\frac{x - x_k}{1 - x_k}, &x > x_k
\end{cases}
, \ x \in [0,1], \ k \in 1,\dots,M,
\end{align}
form a basis for the space of piecewise linear functions with $M$ knots on $[0,1]$, which are $0$ at the origin. We therefore represent each $f_i$ as a linear combination of these basis functions,
\begin{align}
f_i(x_i) = \sum_{k = 1}^{M} \omega_{k,i} \varphi_k(x_i).
\end{align}

The weights $\omega_k$ are the subject of the hierarchical complexity shrinkage procedure, and the structure is very similar to that employed by the grouped LN-CASS prior. The motivation is that, if the $\omega_{k_i}$ are all $0$ for a given, $i$, the covariate has no effect, if we allow $\omega_{1,i}$ to be non-zero, we obtain a linear effect, and if other weights are allowed to be non-zero we obtain a piecewise linear effect. This is the complexity hierarchy we wish to impose.

Thus, the prior on the weights is as follows
\begin{align}
\omega_{1,i}|\lambda_{1,i} & \sim \mathcal{N}(0,(\lambda_{1,i}\tau)^2),\\
\lambda_{1,i} & \sim \textrm{LogitNormal}(\mu_\lambda,\sigma_\lambda), \\
\omega_{k,i}|\lambda_{1,i},\lambda_{k,i} & \sim \mathcal{N}(0,(\lambda_{1,i}\lambda_{k,i}\tau)^2), \ \textrm{for each } k = 2,\dots,M\\
\lambda_{k,i} & \sim \textrm{LogitNormal}(\mu_\lambda,\sigma_\lambda), \ \textrm{for each } k = 2,\dots,M.
\end{align}
Again, by rewriting this in terms of logit transformed normal random variables, we obtain a multivariate normal prior on our parameters of interest.

\subsection{Simulation study details}

Details of the method for generating the synthetic data used in the simulation study (Section II.C.1) are presented here. Table \ref{simstudytable} outlines the ground truth parameters used to generate the data. We drew 100 synthetic covariate vectors from a unit latin hypercube and simulated data from a linear regression with additive Gaussian noise using the parameters generated by the procedure outlined in Table \ref{simstudytable}.

\end{document}